\documentstyle[aps,preprint]{revtex}
\begin{document}
\draft
\preprint{IMSc/94-95/61}
\title{Aspects of Planckian Scattering beyond the Eikonal}
\author{Saurya Das and Parthasarathi
Majumdar\footnote{E-Mail:~saurya,partha@imsc.ernet.in}}
\address{The Institute of Mathematical Sciences, \\ CIT Campus,
Madras - 600 113,  India.}
\maketitle
\begin{abstract}
We discuss an approach to compute two-particle scattering amplitudes for
spinless particles
colliding at Planckian centre-of-mass energies, with increasing momentum
transfer away from
the eikonal limit. For electrically neutral particles, the amplitude exhibits
poles on the
imaginary squared cm energy axis at locations that are distinct from those
appearing in the
eikonal limit. For charged particles, electromagnetic and gravitational effects
remain
decoupled for the eikonal situation as also the leading order (in momentum
transfer, or
equivalently, the impact parameter) correction, but mix non-trivially for
higher orders.
\end{abstract}
\newpage

\section{Introduction}

The efficacy of the shock wave picture \cite{aich} in the computation of
two-particle
scattering amplitudes \cite{thf} for large $s$ (squared centre-of-mass energy)
and small,
fixed $t$ (squared momentum transfer), in the eikonal limit ${s \over t}
\rightarrow
\infty$, is now well-established, both for gravitational and electromagnetic
interactions
[3-8]. The graviton (photon) exchange ladder graphs neatly sum in this
kinematical limit to
reproduce exactly the semiclassical amplitude of the relatively slower test
particle
scattering off the
gravitational (electromagnetic) shock wave due to the ultrarelativistic
`source' particle.
Phenomena beyond this highly restrictive kinematical regime (e.g., for higher
values of
$t$) entail, for their analysis, a calculational scheme for systematic
corrections to the
eikonal. For electrodynamics, with values of $t$ still sub-Planckian, this is
afforded
easily by the usual perturbative formulation of quantum electrodynamics. For
gravity and
electrodynamics of electric {\it and} magnetic charges, the lack of a proper
local quantum
field theory
is a major setback to this programme. On the other hand, a determination of
corrections to
the eikonal is essential to unravel certain features of eikonal scattering
itself, like the
analytic structure (in complex $s$-plane) of the eikonal amplitude \cite{thf},
or the possible
interplay between electromagnetic and gravitational effects for charged
particle
eikonal scattering \cite{dm2}. One approach which has probed the first of these
features
with some success is the one based on reggeized string exchange amplitudes with
subsequent
reduction to the gravitational eikonal limit including the leading order
corrections \cite{amat}. In this letter, we follow a somewhat different
approach
\cite{sing} : the
scattering amplitude is calculated quantum {\it mechanically} by solving the
Klein Gordon
equation of the ultrarelativistic particle in the linearized classical
Schwarzschild
background of the slower `target' particle in the appropriate Lorentz frame.
Recall that
the role of the scattering particles is the opposite to that in the shock wave
picture
\cite{thf} where the slower particle scatters off the shock wave due to the
luminal one.
But this switching allows us to investigate leading corrections to the eikonal.
The
restriction to the linearized Schwarzschild background essentially delineates
the inherent
limitation in our approach vis-a-vis large (e.g. Planckian) momentum transfers;
the latter
situation
does indeed require a full quantum theory of gravity, and is therefore not
immediately
tractable. Admittedly, our approach has been anticipated in analyzing
gravitational eikonal
scattering in earlier work \cite{kab}. Our intention in what follows is to
consider
implications of this `Coulomb scattering' technique {\it beyond the eikonal}.

\section{Purely Gravitational Scattering}

The massless generally covariant Klein Gordon equation for the
ultrarelativistic `test'
particle is given by
\begin{equation}
D_{\mu}D^{\mu} \phi~=~0~.
\end{equation}
In the classical Schwarzschild background of the slow target particle (of mass
$M$ which is
also considered small in comparison with $\sqrt{s}$)
\begin{equation}
ds^2~=~-\left(1-{2GM\over r}\right)dt^2 + {\left(1 - {2GM \over
r}\right)}^{-1}dr^2 + r^2\left(d\theta^2 + \sin^2\theta d\phi^2
\right)~,
\end{equation}
we assume a solution of the Klein Gordon wave function of the form
$$\phi\left(\vec r,t\right) = {f(r) \over r}~ e^{iEt}~
Y_{lm}\left(\theta,\phi\right)~,$$
where $E$ is the energy of the test particle as measured by an asymptotic
observer. On linearizing the Schwarzschild metric, substituting $s=2ME$
and
discarding terms proportional to
$(2GM/r)^2$ or higher powers thereof, we finally obtain the radial part
of the wave equation as:
\begin{equation}
{d^2f(r) \over dr^2} - \left[ {l(l+1) -G^2s^2 \over r^2} - {2
GsE \over r} - E^2 \right] f(r) ~=~0~~. \label{radl}
\end{equation}
Thus, terms with inverse powers in $r$ higher than 2 have been dropped. This
enables us to
solve the resulting eq. (\ref{radl}) without further approximations, while
keeping in mind
that very small (Planck size) impact parameter scattering cannot be probed
thus.

The radial equation (\ref{radl}) above is solved using standard techniques
\cite{lan},
\cite{davy} in terms of hypergeometric functions with well-known asymptotic
properties.
The
scattering amplitude is best expressed in terms of a partial wave expansion, in
view of
the spherical symmetry of the `potential' above,
\begin{equation}
f(\theta)~=~{1 \over 2i \sqrt{s}}\sum_{l=0}^{\infty} (2l+1)
\left[e^{2i\delta_l} -1 \right]
P_l(cos \theta)~ , \label{amp}
\end{equation}
where, the phase shift of the partial wave, characterized by
a fixed angular momentum quantum number $l \gg 1$, is given by
\begin{equation}
\delta_l(s)~=~\arg~\Gamma\left(p_l(s)+1-iGs\right)~, \label{phsft}
\end{equation}
with $p_l(s)$ defined by the relation
\begin{equation}
p_l(s) \left (p_l(s) + 1 \right) \equiv l(l+1) - G^2s^2~~. \label{pel}
\end{equation}
It is not difficult to show from Eqs. (\ref{phsft}) and (\ref{pel}) that, for
fixed $l$,
the phase shift has singularities at cm energies
\begin{equation}
Gs~=~{i \over (2N+1) } \left[ l(l+1)~-~N (N+1) \right] ~~, \label{pol}
\end{equation}
for any non-negative integer $N$. Although still located on the imaginary axis
of the
complex $s$-plane, clearly the locations of these poles are quite distinct from
those
seen in the eikonal limit \cite{thf}, viz., at $Gs~=-iN$. There is also another
distinction: the poles discerned by us are singularities of the phase shift
(for fixed
$l$) and therefore are physically more appealing (i.e., they are most likely
actual
resonances) than the eikonal (large $l$) poles which are not singularities of
the phase
shift \cite{ver}, \cite{amat}. Recall that, for eikonal scattering, $\delta_l
\sim
\log{l}$, but it is perhaps incorrect to suggest that the amplitude has an
$s$-wave pole
because such a low range of $l$ cannot be probed within the eikonal
approximation. In
contrast, while we also cannot probe very low values of $l$, the poles do arise
for
intermediate impact parameter ranges, in the phase shifts themselves. We do not
claim
a full understanding of the origin of these poles, but still feel it useful to
point
out their existence outside the eikonal limit.

The formulas above also permit us to extract the leading order corrections to
the
eikonal limit $l \rightarrow \infty$, by using the asymptotic expansion of the
argument of the gamma function \cite{abrm} in increasing inverse powers of $l$.
We obtain
\begin{equation}
\delta_l~\approx~-Gs \left [ \log{l}- {1 \over 2l} \right] ~+~{(Gs)^3 \over
{2l^2}}
+ O\left( 1\over l^3\right)~. \label{asmp}
\end{equation}
The first term in eq. (\ref{asmp}) obviously corresponds to the eikonal result,
and
the sub-leading corrections have been anticipated from reggeized string
exchange
diagrams \cite{amat}. The leading correction above to the eikonal phase shift
behaves
$Gs/l$. This is
somewhat different from the leading correction as seen in the string theory
based
approach, which is proportional to ${(Gs)^2 \over l^2} \log{s}$. In our quantum
mechanical
approach we do not expect to obtain $\log{s}$ corrections; one needs the
formalism of quantum field theory for that purpose. However, it is a bit
surprising
that a $1/l$ type correction is not obtained in the approach of \cite{amat}.
One
possible explanation could be that the string theory in question is quantized
around a
flat, rather than a Schwarzschild, background and therefore misses this effect.
Even
if this were true, it is not easy to determine the corresponding spacetime
geometry
around the target particle beyond the eikonal limit. That is to say, it remains
to be
seen whether the correction we have found can be interpreted as a contribution
to the
{\it shift} of the appropriate null coordinate found in \cite{aich} which has a
step
function
discontinuity in the other null coordinate, or is it a smearing of the shock
wave
found in the eikonal limit, by exchange of transverse gravitons, as seen in
\cite{amat}.

The asymptotic behaviour observed in eq. (\ref{asmp}) can now be translated
easily to
calculate the scattering amplitude to incorporate the leading order correction;
the
partial wave sum is replaced by the integral over the impact parameter $b
\equiv l/E$,
with the phase shift being replaced by the first two terms in (\ref{asmp}). If,
once
again, the integral is taken between 0 and $\infty$ as in \cite{ver}, it can be
performed exactly, leading to the result
$$ f(s,t) = f^{(0)}(s,t) + f^{(1)} (s,t) + ...,$$
where $f^{(0)}(s,t)$ is the eikonal amplitude. The expressions for these
amplitudes are
\begin{eqnarray}
f^{(0)}&~ =~& {Gs^{3/2} \over 2t} {\Gamma (1-iGs) \over \Gamma
(1+iGs)} {\left( -t \over s \right)}^{iGs} \nonumber \\
f^{(1)}&~ =~& - {Gs \over \sqrt {-t}} {\Gamma ( 1/2 - iGs ) \over
\Gamma ( 1/2 + iGs )} \left( -t \over s\right)^{iGs} ~~.\label{ampl}
\end{eqnarray}
Thus, the eikonal poles are
again manifest at integral values (in Planck units) on the imaginary axis of
the complex
$s$-plane; in
addition, one also observes poles, again originating from the lower limit of
the
integration ($b=0$), at {\it half}-integral values on the imaginary $s$-axis.
As remarked
above, our technique cannot illuminate the really small impact parameter
regime, and thus
these poles are not expected to indicate true resonances because the phase
shift, in the
large l approximation, has no singularities. Therefore, we have very little to
add to the
extant wisdom \cite{ver}, \cite{amat} on the issue of singularities of the
eikonal
amplitude.

\section{Inclusion of Electromagnetism}

Another key issue in Planckian scattering, and one which has not received too
much
attention, is that of mixing of gravitational and electromagnetic effects in
the
eikonal approximation. In the earlier literature \cite{thf}, \cite{dm1}, it was
{\it
assumed} that in the eikonal limit, the gravitational and electromagnetic shock
waves acted quite independently, producing a net phase factor in the wave
function
of the test particle that was a sum of the individual phase factors. Since
generically gravity
couples to everything including electromagnetism, it becomes important to
ascertain
whether the assumed independence of the two interactions in the special
kinematics
of the eikonal limit, really holds. This issue was first addressed in
\cite{dm2}
where heuristic arguments were advanced to show that the assumed decoupling did
indeed take place, thus vindicating results obtained using this crucial
assumption.
The present framework provides a less heuristic avenue to re-examine this
question,
and allows us to establish the earlier conclusions on a sounder footing. In
addition, the decoupling of gravitational and electromagnetic effects is seen
to
persist through the leading order (in inverse powers of the impact parameter)
correction to the eikonal.

As in \cite{dm2}, one begins by considering first the scattering of a (luminal)
neutral test particle off the {\it Reissner-Nordstr\"om} metric due to a static
point
charge. The Klein-Gordon equation of the fast particle can again be written
down by
replacing the spacetime derivatives by generally covariant derivatives
appropriate to the Reissner-Nordstr\"om metric
\begin{equation}
ds^2~=~~-(1-{2GM \over r} + {GQ^2 \over r^2}) dt^2
{}~+~(1-{2GM \over r} + {GQ^2 \over r^2})^{-1} dr^2 ~+~r^2 d \Omega^2~,
\label{rn}~~
\end{equation}
where $d\Omega^2$ is the metric on the unit two-sphere.
Once again, confining ourselves to impact parameters that are large compared to
the
length scale $2GM$ and charges that are of order the electronic charge, the
radial
equation reduces to
\begin{equation}
{d^2f(r) \over dr^2} - \left[ {l(l+1)+2G Q^2 E^2 -G^2s^2 \over r^2} - {2
GsE \over r} - E^2 \right] f(r) ~=~0~~. \label{radlrn}
\end{equation}
The phase shift, for $l \gg 1$ are once again given by eq. (\ref{phsft}),
where,
now
\begin{equation}
p_l(s)\left( p_l(s) + 1 \right)~\equiv~l(l+1)~-~\left(\zeta Gs
\right)^2~~\label{pell}
\end{equation}
with $\zeta^2 \equiv 1-Q^2/2GM^2$. Clearly, $\zeta=1$ is the reduction to the
Schwarzschild case. It is not difficult to show that the phase shift
singularities
now occur not only on the imaginary axis of the complex $s$-plane, but
elsewhere in the
plane as well:
\begin{equation}
Gs~=~\frac i2 \left( {2N+1 \over 1-\zeta^2} \right ) \left \{ -1~+~ \left[
1~-4i
\left( {1-\zeta^2 \over 2N+1} \right ) (Gs)_0 \right ]^{\frac12} \right \}~~,
\label{polrn}
\end{equation}
where, $(Gs)_0$ signifies the location of the poles in the Schwarzschild case,
given
by eq. (\ref{pol}). The $\zeta \rightarrow 1$ limit to the Schwarzschild case
is
again obvious. Apart from reporting the existence of these poles as
singularities of
the phase shift (for intermediate impact parameter ranges), we are unable, at
this
point, to delve deeper into their true origin or full ramification.

One may expand asymptotically the Gamma function in eq. (\ref{phsft}) to
extract
the eikonal limit and the leading order correction; we obtain
\begin{equation}
\delta_l(s)~~=~~-Gs \left[ \log l~-~ {1 \over 2l} \right]~+~\zeta^2
{(Gs)^3 \over 2l^2}~+~ O\left({1 \over l^3 }\right) ~~.
\label{asmp2}
\end{equation}
Clearly, the eikonal term and the leading order correction (the two terms in
the
first pair of square brackets) are completely independent of the charge $Q$ on
the static `target' particle whose gravitational field we have modelled through
the metric of a Reissner-Nordstr\"om back hole. The subleading corrections
(i.e.,
terms of $O(l^{-3})$ or smaller), in contrast,  certainly depend on this
charge. In
other words,
the gravitational effect is completely decoupled from the electromagnetic
effect
for these first two contributions to the phase shift. The mixing that one
expects to
see generically, indeed appears for smaller values of the impact parameter
(smaller $l$). Admittedly, the coefficient of the mixing terms calculated above
is not universal in the sense that one expects corrections to it from
transverse
graviton exchange \cite{amat}; but at least for the first two terms, we expect
our results to be robust.

Further evidence for the decoupling of gravitation and electromagnetism for the
eikonal and leading correction terms, comes
from the scattering of a {\it charged} particle (of charge $Q'$ say) off the
gravitational and electromagnetic field due to the target. This is seen by
generalizing the generally covariant derivatives in the Klein Gordon equation
of the luminal particle, to be $U(1)$ gauge
covariant as well. The radial equation turns out to be a modified version of
eq. (\ref{radlrn}) :
\begin{equation}
{d^2f \over dr^2} - \left[ {l(l+1) + \lambda (Gs)^2 - \alpha'^2 \over
r^2} - {2 \alpha'E \over r} + E^2 \right] f = 0~~\label{radlee}
\end{equation}
where $\alpha'\equiv Gs - QQ'$ and $\lambda$ is defined
by $ \lambda\equiv 1-\zeta^2$.
The asymptotic expansion of the corresponding phase
shift can now be obtained as before with very little extra work. It has the
form
\begin{equation}
\delta_l(s) = -\alpha' \left[ \log l ~-~ \frac{1}{2l} + { \lambda (Gs)^2
- \alpha'^2 \over 2l^2} +  O\left( \frac{1}{l^3}\right) \right]~~.
\label{phsee}
\end{equation}
The replacement $Gs \rightarrow Gs-QQ'$ \cite{thf} to account for the
electromagnetic effects in the eikonal limit, is thus clearly correct within
our
approach. Moreover, such replacement also appears to be equally valid for the
leading
order correction to the eikonal. The mixing between gravity and electromagnetic
effects starts from the `non-universal' $O(l^{-2})$ terms\footnote{The actual
computation of
these terms would be sensitive to the precise manner in which graviton
loop ultraviolet divergences are handled, i.e., on a particular proposal
(model) for a
theory of quantum gravity.} where one expects them to appear in any case.

\section{Conclusions}

While our (semiclassical) method of computing corrections to the eikonal
scattering amplitude
appears
viable, strictly speaking the predictions from this approach are reliable only
for the
leading order correction to the eikonal. The subleading terms within our
approach are
affected nontrivially under true quantum gravitational effects, similar to the
inevitable
necessity of field theoretic quantum electrodynamics for a proper calculation
of the Lamb
shift. The difference here is the lack of an appropriate quantum
`gravidynamics' which can
be reliably used for computation. Since the issue at hand seemingly entails
uncontrollable
ultraviolet behaviour of a local field theoretic formulation of gravity,
starting from
Einsteinian general relativity, the use of string theory to tame these
divergences is
certainly an attractive option. On the other hand, the robustness of the
eikonal amplitude
may indicate certain non-perturbative aspects of spacetime geometry at short
distances
which may not be analyzable in terms of perturbative string theory.

It is satisfying to note that our heuristic analysis on non-mixing of
electromagnetic and gravitational effects for eikonal scattering \cite{dm2} can
indeed be
placed on firmer footing. Likewise, the persistence of this decoupling for the
leading
corrections leads us to infer that these corrections have a similar degree of
universality,
not shared by the higher order effects. The regime of validity of the
semiclassical
approximation for Planckian scattering appears then to have been determined to
a reasonable
degree of accuracy.

Finally, a word about dilaton gravity. The same heuristic arguments which
enable us to
show the decoupling of gravity and electromagnetism in general relativity,
leads to a non-trivial mixing of these interactions even in the eikonal
approximation for
the case of dilaton gravity \cite{dm2}. This also seems to be the case when the
technique
of this paper is applied to dilaton gravity. One is left with the disturbing
possibility
that the inclusion of the dilaton might actually make the eikonal limit
non-existent! We
hope to report on this elsewhere in the near future.

It is a pleasure to thank Prof. V. Singh for suggesting the approach followed
in this
paper to estimate corrections to the eikonal scattering amplitude, and also for
a very
illuminating discussion.

\end{document}